\documentclass[12pt]{article}

\usepackage{graphicx}

\newcommand{\beq}{\begin{equation}}
\newcommand{\eeq}{\end{equation}}
\newcommand{\bea}{\begin{eqnarray}}
\newcommand{\eea}{\end{eqnarray}}

\newcommand{\be}{\begin{equation}}
\newcommand{\ee}{\end{equation}}
\newcommand{\ba}{\begin{array}}
\newcommand{\ea}{\end{array}}

\newcommand{\al}{\alpha}

\newcommand{\de}{\delta}

\newcommand{\ep}{\epsilon}

\newcommand{\la}{\lambda}
\newcommand{\lam}{\lambda}

\newcommand{\om}{\omega}

\newcommand{\La}{\Lambda}

\newcommand{\rar}{\rightarrow}
\newcommand{\non}{\nonumber}
\newcommand{\ts}{\textstyle}

\def\eqs#1{(\ref{#1})}

\font\cmss=cmss10 at 11pt \font\cmsss=cmss8 at 8pt

\def\inbar{\vrule height1.5ex width.4pt depth0pt}
\def\mininbar{\vrule height.75ex width.3pt depth0pt}
\def\cc{\relax\,\hbox{$\mininbar\kern-.2em{\hbox{\rm\tiny C}}$}}
\def\IZ{\relax\ifmmode\mathchoice
{\hbox{\cmss Z\kern-.4em Z}}{\hbox{\cmss Z\kern-.4em Z}}
{\lower.4pt\hbox{\cmsss Z\kern-.4em Z}}
{\lower1.2pt\hbox{\cmsss Z\kern-.4em Z}}\else{\cmss Z\kern-.4em Z}\fi}
\def\IC{\relax\,\hbox{$\inbar\kern-.3em{\rm C}$}}
\def\IR{\relax{\rm I\kern-.18em R}}

\newcommand{\SU}{\mathrm{SU}}

\newcommand{\U}{\mathrm{U}}

\newcommand{\PP}{\mathrm{I}\kern -2pt \mathrm{P}}
\newcommand{\R}{\mathrm{I}\kern -2.5pt \mathrm{R}}
\newcommand{\Z}{\mathsf{Z}\kern -5pt \mathsf{Z}}
\newcommand{\1}{1\kern -3pt \mathrm{l}}

\newcommand{\tr}{{\rm tr}}

\newcommand{\pa}{\partial}

\newcommand{\cO}{\mathcal{O}}
\newcommand{\cN}{\mathcal{N}}
\newcommand{\cF}{\mathcal{F}}

\newcommand{\D}{{\rm d}}
\newcommand{\e}{{\rm e}}

\def\half{ {\textstyle\frac{1}{2}} }

\def\eps{ \epsilon }
\def\Sig{\Sigma}

\newcommand{\tQ}{\tilde{Q}}
\newcommand{\tq}{\tilde{q}}

\def\con{ {\rm const}}
\def\ith{ {i^{\rm th}~{\rm cut}}}
\def\hLa{\hat{\Lambda}}
\def\pre{ \cF(a) }
\def\laSW{ \lambda_{SW} }
\def\gs{ g_{s} }
\def\free{ F_{\rm s}(e, S) }
\def\fre{ F_{\rm s} }
\def\fdisk{ F_{\rm d} (e, S) }
\def\fdis{ F_{\rm d} }
\def\freen{ F^{(n)}_{\rm s}(e, S) }
\def\fdiskn{ F^{(n)}_{\rm d} (e, S) }
\def\freethree{ F^{(3)}_{\rm s}(e, S) }

\newcommand{\fdisktwo}{ F^{(2)}_{\rm d}(e, S) }
\def\Wp{ W^\prime_0}
\def\Weff{ W_{\rm eff}}
\def\tauo{\tau_0} 
\def\gpi{ \gamma_{p,i} }
\def\eij{  e_{ij} }
\def\eik{  e_{ik} }
\def\ejk{  e_{jk} }
\def\eji{  e_{ji} }
\def\eil{  e_{i\ell} }
\def\ekl{  e_{k\ell} }
\def\aij{  a_{ij} }
\def\aik{  a_{ik} }
\def\ajk{  a_{jk} }
\def\aji{  a_{ji} }
\def\ail{  a_{i\ell} }
\def\fiI{  f_{iI} }
\def\fjI{  f_{jI} }
\def\fiJ{  f_{iJ} }
\def\fkI{  f_{kI} }
\def\Li{ L_i}
\def\Lj{ L_j}
\def\Lk{ L_k}
\def\Ll{ L_\ell}

\def\tW{ \tilde{W} }
\def\tZ{ \tilde{Z} }
\def\tS{ \tilde{S} }
\def\tWeff{ \tilde{W}^i_{\rm eff} }
\def\tfree{ \tilde{F}^i_{\rm s} (e, S, \eps) }
\def\tdisk{ \tilde{F}^i_{\rm d}  (e, S, \eps) }
\def\delFhalf{\delta F^i_{\rm d}}
\def\delFnil{\delta F^i_{\rm s}}
\def\delS{\delta S}

\def\poly#1{\tilde{T}(#1)}

\def\sumN{ \sum_{i=1}^{N} }
\def\sumI{ \sum_{I=1}^{N_f} }
\def\sumjN{ \sum_{j=1}^{N} }

\def\prodN{ \prod_{i=1}^{N} }
\def\prodkN{ \prod_{k=1}^{N} }

\def\sumk{ \sum_{k\neq i} }
\def\suml{ \sum_{\ell \neq i,k} }

\def\sphere{ \bigg|_{\rm sphere} }
\def\disk{ \bigg|_{\rm disk} }
\def\vev#1{ \langle {#1} \rangle }
\def\tadpole{ \vev{\tr\, \Psi_{ii} }\big|_{\rm sphere} }
\def\disktadpole{ \vev{\tr\, \Psi_{ii} }\big|_{\rm disk} }
\def\vevS{ \bigg|_{\vev{S}}  }

\begin{document}

\begin{flushright}
BRX-TH-507\\
BOW-PH-127\\
{\tt hep-th/0211254}
\end{flushright}
\vspace{.3in}
\setcounter{footnote}{0}
\stepcounter{table}

\begin{center}

{\Large{\bf\sf Matrix model approach to the {\large $\cN=2$ $\U(N)$} 
gauge theory  \\ with matter in the fundamental representation }}

\vspace{.2in}

Stephen G. Naculich\footnote{Research
supported in part by the NSF under grant PHY-0140281.}$^{,a}$,
Howard J. Schnitzer\footnote{Research
supported in part by the DOE under grant DE--FG02--92ER40706.}$^{,b}$,
and Niclas Wyllard\footnote{Research 
supported by the DOE under grant DE--FG02--92ER40706.\\
{\tt \phantom{aaa} naculich@bowdoin.edu;
schnitzer,wyllard@brandeis.edu}\\}$^{,b}$

\vspace*{0.3in}

$^{a}${\em Department of Physics\\
Bowdoin College, Brunswick, ME 04011}

\vspace{.2in}

$^{b}${\em Martin Fisher School of Physics\\
Brandeis University, Waltham, MA 02454}

\end{center}

\vskip 5mm

\begin{abstract}
We use matrix model technology to study the $\cN=2$ $\U(N)$ gauge 
theory with $N_f$ massive hypermultiplets in the fundamental 
representation.  We perform 
a completely perturbative calculation of the periods $a_i$ 
and the prepotential $\cF(a)$ up to the first instanton level,  
finding agreement with previous results in the literature. We 
also derive the Seiberg-Witten curve 
and differential  
from the large-$M$ solution of the matrix model. 
We show that the two cases $N_f<N$ and $N \le N_f < 2N$ can be 
treated simultaneously.
\end{abstract}

\section{Introduction}
\setcounter{equation}{0}

Dijkgraaf, Vafa, and collaborators have discovered remarkable 
relations between perturbative matrix models and instanton effects 
in supersymmetric gauge 
theories~\cite{Dijkgraaf:2002a}--\cite{Dijkgraaf:2002d}. 
Recently  we  used the new matrix model technology 
to study the $\cN=2$ $\U(N)$ gauge theory \cite{Naculich:2002} 
(ref. \cite{Naculich:2002} also 
contains a more extensive list of references). We 
calculated the prepotential $\mathcal{F}(a)$ and the periods $a_i$ 
perturbatively up to the first 
instanton level. A new ingredient in our calculation was 
a completely perturbative definition of the periods $a_i$ as functions 
of the classical moduli $e_i$.
Our results combined with those of Dijkgraaf and Vafa show that, 
even when the matrix model cannot be completely solved,
a perturbative diagrammatic expansion of the matrix model 
can still be used to obtain all the low-energy non-perturbative 
information of $\cN=2$ gauge theories order-by-order in the 
instanton expansion.

In this paper we study the $\cN=2$ U($N$) gauge 
theory with $N_f$ hypermultiplets transforming in the fundamental 
representation of the gauge group using matrix model techniques. 
Several new features present themselves in this case, making the 
model well worth studying.

In the first part of the paper, we extend the perturbative results 
obtained in~\cite{Naculich:2002} for the $\cN=2$ $\U(N)$ theory to the case 
with $N_f$ fundamental matter hypermultiplets. 
A new feature of the calculation, compared to the one in~\cite{Naculich:2002},
is the appearance of planar diagrams with boundaries \cite{Argurio:2002}.
These contribute, in the diagrammatic expansion of the matrix model,  
to the free energy and superpotential. They also affect the relation between 
the periods $a_i$ and the classical moduli $e_i$. We
compute the periods $a_i$ and prepotential $\cF(a)$ perturbatively
to first order in the instanton expansion, finding agreement 
with earlier results in the literature. This agreement 
is a test of our proposed relation \cite{Naculich:2002} between 
$a_i$ and $e_i$.

In the case of U($N$) with fundamental matter, there is an ambiguity in the 
form of the Seiberg-Witten curve 
\cite{Seiberg:1994}
for $N \le N_f < 2N$ 
\cite{Hanany:1995}--\cite{Krichever:1996}, 
with different forms of the curve
corresponding to different definitions of the classical moduli $e_i$.
These different curves yield slightly different relations between
$a_i$ and $e_i$. 
Our perturbative calculation, which does not start from a curve, 
yields an unambiguous relation between $a_i$ and $e_i$
and therefore implies a particular form the of the Seiberg-Witten
curve, which we show to be
$ y^2 = \prod_{i=1}^N (x-e_i)^2 - f_{N-1}(x)$
where $f_{N-1}(x)$ is an $(N-1)$th order polynomial 
specified in eq.~\eqs{ourcurve}.

In the second part of the paper we derive the form of 
the Seiberg-Witten curve 
and differential  
for the $\cN=2$ $\U(N)$ gauge theory with $N_f$ fundamental 
hypermultiplets, from the large-$M$ saddle-point solution to the matrix 
model, without any additional input. 
The result is consistent with known 
results~\cite{Seiberg:1994}--\cite{Krichever:1996} and also agrees with 
the form of the curve implied by 
the perturbative calculation. Our results give further support 
to the idea that all the low-energy information about the $\cN=2$ theory is 
contained in the matrix model\footnote{The matrix model also knows about 
string theory corrections in the form of curvature 
couplings~\cite{Dijkgraaf:2002c}; 
some such couplings were recently computed~\cite{Klemm:2002} using 
matrix model techniques.}. 
(Very recently some aspects of the relation 
between matrix models and Seiberg-Witten theory 
have been discussed in ref.~\cite{Itoyama:2002}.)

An important question is why the matrix 
model approach to supersymmetric gauge theories   
works and what the scope and limitations of the method are. 
Recently, these questions have been explored and purely 
field-theoretic proofs  
for the correctness of the matrix model approach have been presented 
for the pure $\cN=1$ $\U(N)$ gauge theory with an arbitrary  
polynomial superpotential~\cite{Dijkgraaf:2002e,Cachazo:2002b}.
It would be interesting to extend these 
results to cover the model studied in this paper. 
Also, ref.~\cite{Gopakumar:2002} 
discusses some aspects of the correspondence between 
matrix-model and gauge-theory quantities.

In sec.~2 we set up the perturbative calculation. 
In sec.~3 we calculate $\tau_{ij}$ as a function 
of the classical moduli to 
first order in the instanton expansion. In sec.~4 we extend our proposed 
perturbative definition of the periods $a_i$ to the case when fundamentals 
are present, and use this result to determine
the one-instanton corrections to $a_i$.
In sec.~5 we compute the one-instanton correction to the
prepotential $\cF(a)$. 
When $N_f\ge N$ a certain polynomial appears in the relation 
between $a_i$ and the classical moduli; the role of this polynomial is 
clarified in sec.~6. In sec.~7 we derive the Seiberg-Witten curve
from the large-$M$ saddle point solution to the matrix model. 
In sec.~8 we derive the Seiberg-Witten differential from within 
the matrix model framework. 
We conclude the paper with a summary of our findings.

\section{Perturbative matrix model approach}
\setcounter{equation}{0}

In this section, we describe the perturbative matrix model approach to 
the $\cN=2$ $\U(N)$ gauge theory with matter in the 
fundamental representation,
extending our earlier work~\cite{Naculich:2002}. 
Previous work discussing matter 
in the fundamental representation (focusing mainly on $\cN=1$ theories) 
in the matrix model context can be 
found 
in \cite{Argurio:2002} \cite{McGreevy:2002}--\cite{Feng:2002}. 

In the presence of (massless or massive) $\cN=2$ hypermultiplets transforming 
in the fundamental representation, 
the $\cN=2$ $\U(N)$ gauge theory develops a superpotential 
\be
W_{\mathrm{mat}}(\phi,q,\tq) = 
\sumI \left[ \tq_I \phi\, q^I + m_I \tilde{q}_I q^I  \right] \,,
\ee
written in terms of the $\cN=1$ fields 
$\phi$ (the adjoint scalar in the $\cN=2$ vector multiplet), 
$q^I$ ($I=1,\ldots N_f$) transforming in
the fundamental representation and $\tilde{q}_I$, 
transforming in the conjugate fundamental representation. 
We have suppressed the gauge group indices,
and $m_I$ are the masses of the fundamentals. 

The first step of the matrix model program is 
to break $\cN=2$ supersymmetry to $\cN=1$ 
by adding a tree-level superpotential $W_0(\phi)$ to the gauge theory.
The particular choice of $W_0(\phi)$ relevant to us is the  
one that freezes the moduli to a generic point on the Coulomb branch
of the $\cN=2$ theory:
\beq \label{W}
W_0(\phi) = \al \sum_{\ell=0}^{N} {s_{N-\ell}(e)\over \ell{+}1} 
\tr (\phi^{\ell+1} )
\quad \Rightarrow \quad 
\Wp (x)  = \al \prodN (x-e_i)\,,
\eeq
where $e_i$ are the classical moduli,
$s_m(e)$ is the elementary symmetric polynomial  
\be \label{spoly}
s_m(e) = (-1)^m \sum_{i_1 < i_2 < \cdots < i_m} 
e_{i_1} e_{i_2} \cdots e_{i_m}\,, \qquad\qquad s_0 = 1\,,
\ee
and $\al$ is a parameter that will be taken to zero 
at the end of the calculation, 
restoring $\cN\!=2$ supersymmetry \cite{Cachazo:2002}. 

The next step is to reinterpret the superpotential 
$W(\phi,q,\tilde{q}) = W_0(\phi)+W_{\mathrm{mat}}(\phi,q,\tilde{q})$
as the potential of a chiral 
matrix model~\cite{Dijkgraaf:2002a}-\cite{Dijkgraaf:2002d}, 
which has the partition function 
(denoting the matrix model analogs of 
$\phi$, $q$, and $\tilde{q}$ with capital letters) 
\beq
\label{partition}
Z = {1\over \mathrm{vol}(G)} 
\int \D\Phi \, \D Q^I \D \tQ_I 
\exp \left( - \frac{W(\Phi,Q,\tQ)}{\gs} \right) \,,
\eeq
where the integral is over $M {\times} M$ matrices $\Phi$ 
(which can be taken to be hermitian) 
and $M$-dimensional vectors $Q^I$ and $\tQ_I$.  
In eq.~\eqs{partition}, $G$ is the unbroken matrix model gauge group,
and $\gs$ is a parameter 
that later will be taken to zero as 
$M \to \infty$.
In taking the $M\to\infty$ limit, 
we keep $N_f$ finite (as in ref.~\cite{Bena:2002}); 
our approach thus differs 
from the one in \cite{McGreevy:2002}.
The matrix integral \eqs{partition} is evaluated perturbatively 
about an extremal point $\Phi=\Phi_0$, $Q_0=0$, $\tQ_0=0$
of $W(\Phi,Q,\tQ)$.  We write 
\bea
\label{expand}
\Phi = \Phi_0 + \Psi = \pmatrix{ e_1 \1_{M_1}& 0& \cdots& 0 \cr
                                 0& e_2 \1_{M_2}& \cdots& 0 \cr
                                 \vdots& \vdots& \ddots& \vdots \cr
                                 0& 0& \cdots&  e_N \1_{M_N} }
                     + \pmatrix{ \Psi_{11}& \Psi_{12}& \cdots& \Psi_{1N}  \cr
                                 \Psi_{21}& \Psi_{22}& \cdots& \Psi_{2N}  \cr
                                 \vdots& \vdots& \ddots& \vdots  \cr
                                 \Psi_{N1}& \Psi_{N2} & \cdots&  \Psi_{NN} }\,,
\eea
where $\sum_i M_i = M$, and $\Psi_{ij}$ is an $M_i \,{\times} M_j$ matrix.
This choice breaks the $\U(M)$ symmetry to $G = \prod_{i=1}^N  U(M_i)$. 

The connected diagrams of the perturbative expansion of $Z$ may be organized,
using the standard double-line notation, 
in a topological expansion characterized by the Euler characteristic $\chi$ of 
the surface in which the diagram is  
embedded \cite{tHooft:1974}
\beq
Z =\exp \left(   \sum_{\chi \le 2} \gs^{-\chi} F_{\chi} (e, S) \right)
\quad {\rm where~} \quad S_i \equiv \gs M_i\,,
\eeq
where $\chi = 2 {-} 2g {-} h$ with $g$ the genus 
(number of handles) and $h$ the number of holes. 
When evaluating the matrix integral in 
the $M_i \to \infty$, $\gs \to 0$  limit, with $S_i$ held fixed,
the leading contribution comes from the planar diagrams 
that can be drawn on the sphere ($\chi=2$),
\beq
\free \equiv F_{\chi=2} (e,S)= \gs^2 \log  Z \sphere
\eeq 
As discussed in \cite{Argurio:2002},
the presence of the $Q^I$, $\tQ_I$'s 
leads to the introduction of surfaces with boundaries
in the topological  expansion. 
The leading boundary contribution comes from 
surfaces with one boundary (disks), 
obtained from the sphere by cutting out one hole,
and having $\chi=1$,
\beq
\fdisk \equiv F_{\chi=1} (e,S) =  \gs \log  Z \disk
\eeq

It was shown in \cite{Naculich:2002} (generalizing the result 
in~\cite{Dijkgraaf:2002d} for $\U(2)$) that when one expands $W_0(\Phi)$ 
(\ref{W})
to quadratic order in $\Psi$, the coefficients of  
$\tr (\Psi_{ij} \Psi_{ji})$ vanish when $i{\neq} j$. Hence   
the off-diagonal matrices $\Psi_{ij}$ are zero modes, and correspond
to pure gauge degrees of freedom.  
As in ref.~\cite{Dijkgraaf:2002d,Naculich:2002}, 
we fix the gauge $\Psi_{ij}=0$ ($i{\neq} j$) and introduce 
Grassmann-odd ghost matrices $B$ and $C$ 
with action 
\beq
\tr \left(  B [\Phi, C] \right) = 
 \sumN \sum_{j \neq i} (e_i - e_j) \tr (B_{ji} C_{ij})
+\sumN \sum_{j \neq i} 
\tr (B_{ji} \Psi_{ii} C_{ij} - B_{ji} C_{ij} \Psi_{jj})\,.
\eeq
In the $\Psi_{ij}=0$ ($i{\neq} j$) gauge $W_0(\Phi)$ 
becomes \cite{Naculich:2002}
\beq \label{Wexp}
W_0(\Phi)  =  \sumN   M_i W_0(e_i) + {\al} 
\sumN  {R_i \over 2} \tr(\Psi_{ii}^2) +  
{\al} \sumN \sum_{p=3}^N  {\gpi \over p} \tr(\Psi_{ii}^p)\,,
\ee
where $R_i  =  \prod_{j\neq i} \eij$ with $\eij = e_i - e_j$, and 
\beq
\gpi = {1\over (p-1)! }
\left[ \left( \pa \over \pa x\right)^{p-1} \prodkN  (x-e_k) \right] 
\Bigg|_{x=e_i}
\eeq

Writing $Q^I = (Q^I_1, Q_2^I ,... , Q_N^I)^{T}$, 
where $Q^I_i$ is an $M_i$-dimensional vector and similarly for 
$\tilde{Q}_I$, and expanding $W_{\rm mat}(\Phi,Q^I,\tQ_I)$ around the 
vacuum (\ref{expand}) one finds (using the $\Psi_{ij} = 0$ ($i{\neq} j$) gauge)
\be
W_{\mathrm{mat}}(\Phi,Q,\tQ) = 
\sumN \sumI \left[ (e_i + m_I) \tilde{Q}_{iI} Q_i^I  +  
\tilde{Q}_{iI} \Psi_{ii} Q_i^I \right]\,.
\ee
Collecting the above results, 
the partition function is given by the gauge-fixed integral
\beq
\label{freeenergy}
Z_{\rm g.f.} ={1\over \mathrm{vol}(G)}  
\exp \left(- {1\over \gs} \sumN   M_i W_0(e_i) \right)
 \int \D\Psi_{ii}\, \D B_{ij}\, \D C_{ij} \, \D Q^I \D\tQ_I 
\e^{I_{\rm quad} + I_{\rm int}} 
\eeq
where the quadratic part of the action is
\beq \label{Iprop}
I_{\rm quad}  =  - {\al\over\gs} \sumN  {R_i\over 2} \tr(\Psi_{ii}^2) 
- \sumN \sum_{j\neq i} \eij \tr (B_{ji} C_{ij} ) 
- {1\over \gs} \sumN \sumI (e_i + m_I) \tilde{Q}_{iI} Q_i^I \,,
\eeq
and the interaction terms are
\beq 
\label{Iint}
I_{\rm int} 
=  - {\al\over\gs} \sumN \sum_{p=3}^N  {\gpi \over p} \tr(\Psi_{ii}^p)
-  \sumN \sum_{j \neq i} 
\tr (B_{ji} \Psi_{ii} C_{ij} - B_{ji} C_{ij} \Psi_{jj}) -
{1\over \gs} \sumN \sumI \tilde{Q}_{iI} \Psi_{ii} Q_i^I \,.
\eeq
The propagators for the various fields can be read off 
from eq.~\eqs{Iprop} and the vertices from eq.~\eqs{Iint}.
Each ghost loop will acquire an additional factor of 
$(-2)$~\cite{Dijkgraaf:2002d}.

\section{Perturbative calculation of $\tau_{ij}(e)$}
\setcounter{equation}{0}

The integral over the part of the quadratic action (\ref{Iprop}) 
involving $\Psi_{ii}$, $B_{ij}$, and $C_{ij}$ can be explicitly
performed \cite{Naculich:2002}; including also the 
classical piece one finds
(up to an $e_i$-independent quadratic monomial in the $S_i$'s) 
\beq
\label{freeenergytwo}
\free = 
- \sumN S_i W_0(e_i) 
+ \half \sumN S_i^2 \log \left( S_i\over \alpha R_i \hLa^2 \right) 
+ \sumN \sum_{j\neq i}  S_i S_j \log\left(\eij \over \hLa \right) + 
\sum_{n \ge 3} \freen
\eeq
As in \cite{Dijkgraaf:2002d}, we have included in eq.~\eqs{freeenergytwo}
a contribution $ - \left( \sumN S_i \right)^2 \log \hLa$
that reflects the ambiguity in the cut-off of the full $\U(M)$
gauge group. 
(A similar contribution is included in (\ref{Fdisktwo}) below.)
The term $\freen$ is an $n$th order polynomial in $S_i$
arising from planar loop diagrams built from the interaction
vertices \cite{Dijkgraaf:2002c}.
The contribution to $\free$ cubic in $S_i$ was computed 
in~\cite{Naculich:2002} with the result:
\bea
\label{freeloop}
\alpha \freethree &=& 
({\ts \frac{1}{2} } + {\ts \frac{1}{6} } ) 
\sum_i  \frac{S_i^3}{R_i} \left( \sumk {1 \over \eik} \right)^2
- {\ts \frac{1}{4} }
\sum_i \frac{S_i^3}{R_i} \sumk \suml {1 \over \eik \eil} \non\\ 
&& 
- 2 \sum_i \sumk \frac{S_i^2 S_k}{R_i \eik} \sum_{\ell \neq i} \frac{1}{\eil}  
+ 2 \sum_i \sumk \sum_{\ell \neq i} \frac{S_iS_kS_\ell}{R_i \eik \eil}
-   \sum_i \sumk  \frac{S_i^2 S_k}{R_i \eik^2}\,.
\eea

Next we turn to the contribution of the fundamentals $Q$, $\tQ$
to the matrix model free energy.
Since these involve quark loops, they only contribute to
the disk-level part of the free energy.
The integral over the quadratic part of $W_{\mathrm{mat}}$ 
gives\footnote{Note that there are no $M_i \log M_i$ terms in the expansion 
of $1/\mathrm{vol}(G)$ \cite{Ooguri:2002}.} 
\be
\int \prodN \prod_{I=1}^{N_f} \D Q^I_i \D \tilde{Q}_{iI} 
\exp \left( -\frac{1}{\gs}(e_i + m_I) \tilde{Q}_{iI} Q_i^I \right) = 
\exp \left( - \sumN \sumI M_i 
\log \frac{(e_i + m_I)}{\gs} \right)
\ee
which yields
(up to an $e_i$-independent part linear in $S_i$)
the first term of
\be 
\label{Fdisktwo}
\fdisk = -  \, \sumN \sumI S_i \log\frac{(e_i + m_I)}{\hLa}
+ \sum_{n \ge 2} \fdiskn
\eeq
Here $\fdiskn$ is an $n$th order polynomial in $S_i$
arising from planar disk diagrams 
built from the interaction vertices.
To obtain the $\cO(S^2)$ contribution 
to $\fdisk$, 
we need to evaluate the diagrams displayed in figure 1.

\bigskip \medskip

\begin{figure}[h]
\begin{center}
 \includegraphics{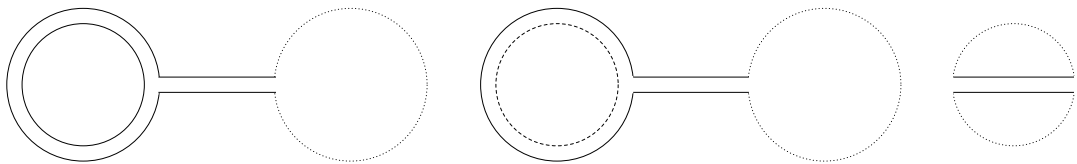}  \\[-3.5cm]
{\bf \caption{ \small \rm Disk diagrams contributing 
to $\fdisk$ at order $\cO(S^2)$. Solid double lines refer to $\Psi_{ii}$ 
propagators, solid-plus-dashed double lines refer to ghost propagators, 
and single dotted lines correspond to the propagator for the $Q$'s. }}
\end{center}
\end{figure}

One might also consider diagrams drawn on surfaces with additional holes. 
One example is a ``dumb-bell'' diagram as in figure 1, but with 
quark propagators at both ends. Such a diagram corresponds to 
 a sphere with two holes, the dotted lines encircling each of the two holes. 
However, such a surface has $\chi=0$ and the diagram is therefore suppressed
by a factor of $\gs$ relative to the
$\chi = 1$ disk contribution
in the $\gs \rar 0$, $M_i \rar \infty$ limit.

The above diagrams lead to:
\be 
\label{diskloop}
\al \fdisktwo = \sumI  
\left[
\sum_i {S^2_i \over R_i \fiI} 
\sum_{j\neq i} \frac{1}{\eij}
-2 \sum_i \sum_{j\neq i} \frac{S_iS_j}{R_i \eij \fiI } 
+ \half \sum_i
\frac{S^2_i}{R_i \fiI^2}
\right]\,,
\ee
where $\fiI = e_i + m_I$.

To relate the matrix model and its free energy
to the $\cN=2$ U($N$) gauge theory (with $N_f$ hypermultiplets in the 
fundamental representation of the gauge group) 
broken to $\prod_i \U(N_i)$, 
one introduces 
\cite{Dijkgraaf:2002a}--\cite{Dijkgraaf:2002c} \cite{Cachazo:2001}
\cite{Argurio:2002}
\beq
\label{Weffdef}
\Weff (e, S) 
= - \sumN  N_i {\partial \free \over \partial S_i} - \fdisk
  +  2 \pi i \tauo \sumN  S_i\,,
\eeq
where $\tauo = \tau(\La_0)$ is the gauge coupling 
of the U$(N)$ theory at some scale $\La_0$.
Since we are breaking $\U(N)$ to $\U(1)^N$,
we set  $N_i=1$ for $i=1,\ldots,N$.
It was conjectured in ref.~\cite{Argurio:2002}
that the disk-level part of the free energy $\fdisk$
contributes to $\Weff$
without any derivatives acting on it. We will find further 
support for this claim. 
Next, one extremizes the effective superpotential with respect to $S_i$ 
to obtain $\vev{S_i}$:
\beq \label{Wextreme}
{\partial \Weff (e, S)  \over \partial S_i} \bigg|_{S_j = \vev{S_j} } =0\,.
\eeq
Finally, 
\beq
\label{matrixtau}
\tau_{ij} (e) = {1\over 2\pi i} 
{\partial^2 \free \over \partial S_i \partial S_j}
\bigg|_{S_i = \vev{S_i} } 
\eeq
yields the couplings of the unbroken U$(1)^N$ factors of the gauge theory,
as a function of $e_i$. 
Note that although both the Seiberg-Witten formula 
$\tau_{ij}(a)=\frac{\pa^2 \cF(a)}{\pa a_i \pa a_j}$ 
and \eqs{matrixtau} 
refer to the same quantity 
(the period matrix of the Seiberg-Witten curve $\Sig$),
they are expressed in terms of different parameters
on the moduli space ($a_i$ vs. $e_i$).

Above, we have evaluated $\free$ to cubic order in $S_i$ and 
$\fdisk$ to quadratic order in $S_i$,
which will be sufficient to obtain $\tau_{ij}(e)$ 
to one-instanton accuracy.
Inserting the results eq.~\eqs{freeenergytwo}, \eqs{freeloop},
\eqs{Fdisktwo}, and \eqs{diskloop}  in eq.~\eqs{Weffdef}, 
we obtain
\bea
\label{effective}
&&\!\!\!\!\!\!\Weff (e,S) =  
\sum_i W_0(e_i) 
- \sum_i  S_i \log \left( S_i \over \alpha R_i \hLa^2 \right) 
- 2 \sum_i \sumk  S_k \log \left( \eik\over \hLa \right)    
+  \, \sum_i \sumI S_i \log \left( \fiI \over \hLa \right)
\non\\
&& 
- {1\over \al} \Bigg[ 
\sum_i \sumk \suml \left( 
- {3 S_i^2 \over 4 R_i \eik \eil} + {2 S_kS_\ell \over R_i \eik\eil} 
\right)
+ \sum_i \sumk \left(
- \frac{S_i^2}{R_i\eik^2}  
-  \frac{2 S_i S_k}{R_i \eik^2 } 
+ \frac{2 S_i^2}{R_k \eik^2}   \right)
\non\\
&&
+ \sumI  \sum_i 
\left( 
- {S^2_i \over R_i \fiI} \sumk {1\over \eik} 
+ \sumk {2 S_iS_k \over R_i \eik \fiI }  
- {S^2_i \over 2R_i \fiI^2}
\right)
\Bigg]
+  \left( 2\pi i \tauo + {\rm const}\right) \sum_i S_i \,.
\eea
The extrema $\vev{S_i}$ are obtained from  (\ref{Wextreme}), 
and can be evaluated in an expansion in $\La$ 
\bea
\label{Svev}
\vev{S_i} &=& 
{\alpha \Li \over R_i}  \La^{2N-N_f} 
+ {\alpha \Li \over R_i}  \La^{4N-2N_f} 
\Bigg[
\sumk \suml  \left(
{3 \Li \over 2 R_i^2 \eik\eil} + {4 \Ll  \over R_k R_\ell \eik \ekl} 
\right)
\non\\
&& 
+ \sumk \left(
\frac{2 \Li}{R_i^2\eik^2} 
- \frac{4\Li}{R_i R_k \eik^2}
+ \frac{2 \Lk }{R_i R_k \eik^2}
+ \frac{2 \Lk}{R_k^2\eik^2}
\right)
- \sumI {\Li  \over R_i^2\fiI^2 }
 \non\\
&&
+ \sumI \sumk \left(
- {2 \Li \over R_i^2\eik\fiI}
+ {2 \Lk \over R_i R_k \eik\fiI}
- {2 \Lk \over  R_k^2 \eik\fkI}
\right)
\Bigg] + \cO(\La^{6N-3N_f})
\eea
where $L_i = \prod_{I=1}^{N_f} (e_i + m_I)$,
and various constants as well as $\tauo$  
have been absorbed
into a redefinition of the cut-off, 
$\La = {\rm const}\, {\times} \hLa \, \e^{\pi i \tauo/N}$.

Although we are primarily interested in the $\cN=2$ limit in
this paper, 
the $\cN=1$ effective superpotential may be easily 
computed by substituting eq.~\eqs{Svev} into eq.~\eqs{effective}. 
In the case $N_f \ge N$ one has to proceed with care, 
see~\cite{McGreevy:2002,Bena:2002,Demasure:2002} for further details.

Below we will make repeated use of the identity
\be 
\label{id}
\sumk {L_k \over R_k \eik} = - {L_i\over R_i} \sumk {1\over \eik} 
+ \frac{L_i}{R_i}\sum_{I} {1\over \fiI} - \poly{e_i}
\eeq
which can be derived by taking the $z \to e_i$ limit of both sides of
\beq
\label{polydef}
\frac{ \prod_{I=1}^{N_f} (z+m_I) }{ \prod_{k=1}^N (z-e_k) }  
- {L_i \over R_i (z-e_i)} = \sum_{k \neq i} {L_k \over R_k (z-e_k)} 
+       \poly{z}\,.
\eeq
where the polynomial $\poly{z} = \sum_{k=0}^{N_f-N} \tilde{t}_k z^{N_f-N -k}$ 
is the positive part of the Laurent expansion 
of ${ \prod_{I=1}^{N_f} (z+m_I) }/{ \prod_{k=1}^N (z-e_k) }$ 
and is only non-zero when $N_f\ge N$.
More explicitly, the coefficients $\tilde{t}_k$ are exactly as in 
eqs. (2.4) and (2.5) of ref.~\cite{D'Hoker:1996}; 
note that our $e_i$ are the same as their $\bar{a}_i$.

We can now evaluate 
\beq
\tau_{ij}(e)  = {1\over 2\pi i} 
{\partial^2 \free \over \partial S_i \partial S_j}
\bigg|_{S_i = \vev{S_i} } 
= \tau_{ij}^{\rm pert}(e) + \sum_{d=1}^\infty \La^{(2N-N_f)d} 
\tau_{ij}^{(d)}(e) \,.
\eeq
The perturbative contribution (up to an additive constant) is
\beq
\label{taupert_of_e}
2\pi i \tau^{\rm pert}_{ij} (e)
=\delta_{ij} 
\Bigg[- \sumk \log \left( \eik \over \La \right)^2 
+ \sumI \log \left( \fiI \over \La \right)
\Bigg]
+ (1 - \delta_{ij}) 
\Bigg[ \log \left( \eij \over \La \right)^2 \Bigg]\,.
\eeq
Using the identity (\ref{id}) one obtains the one-instanton contribution
\bea 
\label{tau1_of_e}
2 \pi i \tau^{(1)}_{ij} (e)\!\!&=&\!\!
\de_{ij} \Bigg[ 
\sumk \suml  \left( 
{8 \Li \over R_i^2 \eik\eil} - {4 \Lk \over R_k^2  \eik\ekl}
\right)
+ \sumk \left(
{10 L_i \over R_i^2 \eik^2} + {10 \Lk \over R_k^2 \eik^2} 
+ {4 \poly{e_i} \over R_i \eik} - {4 \poly{e_k} \over R_k \eik}
\right)
\non\\
&& \;\;\;
+\sumI \left( 
  {\Li \over R_i^2\fiI^2 } 
+ \sum_{J \neq I} {2 \Li \over R_i^2\fiI \fiJ }
- \sumk {8 \Li \over R_i^2 \eik\fiI }
+ \sumk { 2 \Lk \over R_k^2 \eik\fkI } 
- {2 \poly{e_i} \over R_i \fiI }
\right)
\Bigg]
\non\\
&+&(1-\de_{ij})
\Bigg[ 
\sum_{k\neq i,j} \left( -  {8 L_i \over R_i^2 \eij \eik}
			-  {8 L_j \over R_j^2 \eji \ejk} 
                        +  {4 \Lk \over R_k^2  \eik \ejk} \right)
- {10\Li \over R_i^2\eij^2} - {10\Lj \over R_j^2\eij^2}
\non\\
&& \qquad \qquad \;\;
\sumI  \left( {4\Li \over R_i^2 \eij\fiI}
+{4\Lj \over R_j^2 \eji\fjI} \right)
-{4\poly{e_i} \over R_i \eij}
-{4\poly{e_j} \over R_j \eji}
\Bigg]
\eea
to the gauge coupling matrix. Finally,  we take the 
limit $\alpha \to 0$ to restore $\cN=2$ supersymmetry,
but this has no effect on $\tau_{ij}$, which is independent of $\alpha$.

\section{Perturbative determination of $a_i$}
\setcounter{equation}{0}

If we are to use the matrix model 
results \eqs{taupert_of_e}, \eqs{tau1_of_e} to determine
the  $\cN=2$ prepotential $\pre$, 
we must first express $\tau_{ij}$ in terms of the periods $a_i$.
In \cite{Naculich:2002} we proposed a  
definition of $a_i$ within the context of the perturbation expansion of 
the matrix model, without
referring to the Seiberg-Witten curve or 
differential\footnote{For 
the model studied in this paper the Seiberg-Witten curve is 
known~\cite{Seiberg:1994}--\cite{Krichever:1996}  
and the relationship 
between $a_i$ and $e_i$ is straightforwardly 
obtained \cite{D'Hoker:1996} from the $A_i$-period integral. 
However, our goal in this section is to determine $a_i$ using only the 
matrix model perturbation expansion.}.
We argued in \cite{Naculich:2002} 
that $a_i$ can be determined perturbatively via 
\beq
\label{ai} 
a_i = 
{ \pa \tWeff(e,\vev{\tS},\eps) \over \pa \eps } \bigg|_{\eps \to 0}
\eeq
where $\tWeff(e,S,\eps)$ is the effective superpotential that one obtains
by considering the matrix model with action
$\tW^i (\Phi,Q,\tQ) =  W (\Phi,Q,\tQ) + \eps \, \tr_i \, \Phi $. 
Here the trace is only 
over the $i$th block. For motivations for this proposal 
we refer the reader to~\cite{Naculich:2002}. 
In the present case, it is sufficient to consider 
\bea
\label{Ztilde}
\tZ^i
&=& {1\over \mathrm{vol}(G)} 
\int  \D\Phi  \exp 
\left( - \, {1\over \gs} 
\left[ W(\Phi,Q,\tQ) + \eps\, \tr_i \, \Phi \right] \right)  \non \\
&=& 
\exp \left(   {1\over \gs^2} \tfree + \frac{1}{\gs} \tdisk + \ldots \right)\,.
\eea
Writing $\tfree = \free + \eps \delFnil$ 
and similarly for $\tdisk$, and observing that to first order in $\ep$
\beq
\tZ^i = Z + 
{1\over \mathrm{vol}(G)} 
\int  \D\Phi  
\left[ - \,  {\ep \over \gs} \right] \tr_i \, \Phi
\exp \left( -  \, {W(\Phi,Q,\tQ) \over \gs} \right)\,,
\eeq
one finds 
$\delFnil = -\, \gs \vev{\tr_i\,\Phi}\big|_{\rm sphere}$, 
where 
$\vev{\tr_i\,\Phi}\big|_{\rm sphere}$ 
is obtained by calculating 
all connected one-point functions at sphere-level  
in the matrix model with action $W(\Phi,Q,\tQ)$. 
Similarly, 
$\delFhalf = - \vev{\tr_i\,\Phi}\big|_{\rm disk}$ where 
$\vev{\tr_i \, \Phi}\big|_{\rm disk}$ 
is obtained by computing all connected one-point functions at disk-level.

Now the effective potential for the matrix integral \eqs{Ztilde} is
\bea
\label{tWeff}
\tWeff (e,S, \eps) &=&  - \sumjN  N_j {\partial \tfree \over \partial S_j} 
 - \tdisk
+  2 \pi i \tauo \sumjN  S_j \non \\
&=&  \Weff (e,S) - \eps \left[ 
\sumjN N_j {\pa \over \pa S_j} \delFnil + 
\delFhalf \right] \,.
\eea
Extremizing $\tWeff (e,S, \eps)$ with respect to $S$ gives
$\vev{\tS_i} = \vev{S_i} + \eps \delS_i + \cO(\eps^2)$.
Substituting $\vev{\tS}$ into eq.~\eqs{tWeff}, one obtains
\beq
 \tWeff (e,\vev{\tS}, \eps) 
=\Weff (e,\vev{S}) + \eps \sumjN \delS_j {\pa \Weff  \over \pa S_i} \vevS
- \eps \left[ \sumjN N_j {\pa \over \pa S_j} \delFnil 
+ \delFhalf \right]\vevS + \cO(\eps^2)
\eeq
The second term vanishes by the definition of $\vev{S}$.
Finally, using eq.~\eqs{ai}, one obtains
\beq
a_i = - \left[ \sumjN N_j 
{\pa \over \pa S_j} \delFnil  + \delFhalf \right]  \vevS
\eeq
Considering a generic point in moduli space, 
where U$(N) \to \U(1)^N$ (so that $N_i = 1$) and 
expanding $\Phi$ around the vacuum \eqs{expand},
$\tr_i\, \Phi = M_i e_i + \tr(\Psi_{ii}) $, we find
\beq
\label{amat}
a_i =  e_i + \left[ \sumjN {\pa \over \pa S_j} \gs \tadpole 
+ \disktadpole \right] \vevS
\eeq
where $\tadpole$ is obtained by calculating, using 
the matrix model (\ref{freeenergy}), 
all connected planar tadpole diagrams 
with an external $\Psi_{ii}$ leg that can be drawn on a sphere, and 
$\disktadpole$ is obtained by computing 
all connected planar tadpole diagrams 
with an external $\Psi_{ii}$ leg at disk-level in the topological expansion.  

It should be emphasized that (\ref{amat}) offers a 
completely perturbative means of obtaining the relation 
between $a_i$ and $e_i$, which does not require 
knowledge of the Seiberg-Witten curve or differential.

We will now evaluate eq.~(\ref{amat}) for the case 
of the $\cN=2$ $\U(N)$ gauge theory with $N_f$ fundamental
hypermultiplets.  
The relevant tadpole diagrams 
contributing to first order in the instanton expansion
are displayed in figure 2.

\bigskip\medskip

\begin{figure}[h]
\begin{center}
 \includegraphics{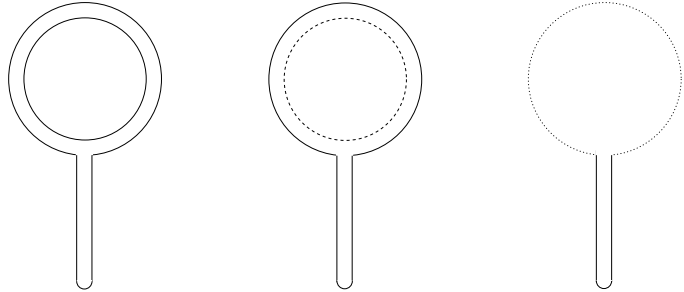}  \\[-2cm]
{\bf \caption{\small \rm Tadpole diagrams contributing to the 
one-instanton contribution to $a_i$.}}
\end{center}
\end{figure}
The first two diagrams contribute to $\tadpole$. These were evaluated 
in \cite{Naculich:2002} with the result
\be
\tadpole = 
{1 \over \al \gs} 
\sum_{j\neq i}
\left[-\frac{S_i^2 }{R_i \eij}
+ 2\frac{S_i S_j}{R_i \eij} \right]\,.
\ee
The third diagram in figure 2 contributes to $\disktadpole$. By using 
the Feynman rules derived from the action \eqs{freeenergy} one finds
\be
\disktadpole = -\, {S_i \over \al R_i}\sumI \frac{1}{\fiI}\,.
\ee
Inserting the above results into eq.~\eqs{amat},
evaluating the resulting expression using eq.~\eqs{Svev},
and using the identity \eqs{id},
one finds 
\be
\label{ae}
a_i = e_i +  \La^{2N-N_f} 
\left( -  {2 \Li \over R_i^2}\sum_{j\neq i}\frac{1}{\eij} 
       +  {\Li \over R_i^2 } \sumI \frac{1}{\fiI}
       -  {2 \poly{e_i}  \over R_i}
\right)
+ \cO(\La^{4N -2N_f})\,.
\ee

The relation between $a_i$ and $e_i$ that we have just
derived agrees precisely, at the one-instanton level, 
with eq. (3.10) of ref.~\cite{D'Hoker:1996}, 
provided that the polynomial $T(x)$ in their expression is 
set equal to $\half \poly{x}$. 
We will discuss the implications of this result in section \ref{sT}. 

\setcounter{equation}{0}
\section{Perturbative calculation of $\tau_{ij}(a)$ and $\cF(a)$}

Now that we have determined the relation between $a_i$ and
$e_i$, we can rewrite $\tau_{ij}(e)$ in terms of $a_i$,
and from that determine the form of the prepotential $\cF(a)$ 
to one-instanton accuracy. Equation \eqs{ae} implies that
\bea
\log \eij  &=& \log \aij
+ \La^{2N-N_f} \bigg[
\sum_{k\neq i,j} \left( {2 \Li \over R_i^2 \eij \eik}
                       +{2 \Lj \over R_j^2 \eji \ejk} \right)
+{2 \Li \over R_i^2\eij^2}
+{2 \Lj \over R_j^2\eij^2}  \non\\
&& -\sumI \left( {\Li \over R_i^2\eij\fiI }
            + {\Lj \over R_j^2\eji\fjI } \right)
+ {2 \poly{e_i} \over R_i \eij}
+ {2 \poly{e_j} \over R_j \eji}
\bigg] \,,
\eea
where $\aij= a_i - a_j$,
and
\beq
\log \fiI  = \log  (a_i+m_I)
+ \La^{2N-N_f} \left[
\sumk {2 \Li \over R_i^2 \eik\fiI }
-  {\Li \over R_i^2\fiI } \sum_{J=1}^{N_f} \frac{1}{\fiJ}
+ {2 \poly{e_i} \over R_i\fiI }
\right] \,.
\eeq
We can now re-express $\tau_{ij}$ (\ref{tau1_of_e})
in terms of $a_i$
\beq
\label{tau_of_a}
\tau_{ij} (a)  =
\tau_{ij}^{\rm pert}(a) + \sum_{d=1}^\infty \La^{(2N-N_f)d} 
\tau_{ij}^{(d)}(a)\,,
\eeq
where the perturbative contribution is (up to additive constants) 
\beq \label{taupert_of_a}
2\pi i \tau^{\rm pert}_{ij} (a)
=
\delta_{ij} 
\Bigg[ - \sumk \log \left( a_i - a_k  \over \La \right)^2 
+ \sumI \log \left( a_i + m_I \over \La \right) \Bigg] 
+ (1 - \delta_{ij}) 
\Bigg[ \log \left( a_i - a_j  \over \La \right)^2 \Bigg] 
\eeq
and the one-instanton contribution is
\bea \label{tauone_of_a}
2 \pi i \tau^{(1)}_{ij} (a) &=&
\de_{ij} \Bigg[ 
\sumk \left( {4 \Li \over R_i^2} \suml {1\over \aik\ail}
+ {6\Li \over R_i^2 \aik^2}
+ {6\Lk \over R_k^2 \aik^2}  \right) \non \\
&& \;\;\; +\sumI \left( 
- \sumk {4 \Li \over R_i^2 \aik\fiI }
+ \sum_{J \neq I} {\Li \over R_i^2\fiI \fiJ }
\right)
\Bigg]  \\
&&+ (1-\de_{ij})
\Bigg[ 
\sum_{k\neq i,j} 
\left( - {4 \Li \over R_i^2 \aij \aik}
       - {4 \Lj \over R_j^2 \aji \ajk}
       + {4 \Lk \over R_k^2  \aik \ajk} \right)
- {6\Li \over R_i^2\aij^2}
- {6\Lj \over R_j^2\aij^2}   \non \\
&& \qquad \qquad \; + \sumI \left( {2 \Li \over R_i^2 \eij\fiI }
             + {2 \Lj \over R_j^2 \eji\fjI } \right)
\Bigg]\,, \non
\eea
where now $R_i = \prod_{j \neq i} (a_i - a_j)$
and $\fiI = a_i + m_I$.
Observe that all the $\poly{x}$ terms cancel out in the
final expression for $\tau_{ij}(a)$.
As will be discussed in more detail in the next section, $\poly{x}$
can be absorbed into 
a redefinition of the $e_i$ \cite{D'Hoker:1996}.  
Since $\tau_{ij}(a)$ is independent of $e_i$
it should be insensitive to this redefinition, 
and therefore to the form of $\poly{x}$.

Finally, it is readily verified that 
(\ref{taupert_of_a}), (\ref{tauone_of_a}) can be written as 
$ \tau_{ij} = {\pa^2 \pre / \pa a_i \pa a_j}$
with (up to a quadratic polynomial)
\bea
2\pi i\pre
&=& -{\ts \frac{1}{4}}\sum_i \sum_{j\neq i} (a_i-a_j)^2 
\log \left(a_i-a_j \over \La \right)^2 
  +{\ts \frac{1}{4}}\sum_i \sumI (a_i+m_I)^2 
\log \left(a_i+m_I \over \La \right)^2  \non\\
&& + \La^{2N-N_f}  \sum_i \prod_{j \neq i} 
\prod_{I=1}^{N_f} {(a_i + m_I) \over (a_i-a_j)^2}
+ \cO(\La^{4N-2N_f})\,.
\eea
This precisely agrees 
with the result obtained in eq. (4.34) of ref.~\cite{D'Hoker:1996}.

To conclude, we have shown that a completely perturbative matrix model 
calculation, 
which does not use the Seiberg-Witten curve or differential,
gives the correct result for the prepotential to first order 
in the instanton expansion for the $\U(N)$ gauge theory with $N_f$ 
fundamentals.
Higher-instanton corrections to the prepotential 
may be obtained by higher-loop contributions to the matrix 
model free energy and tadpole diagrams.

\setcounter{equation}{0}
\section{The meaning of $\poly{x}$} \label{sT}

In ref.~\cite{D'Hoker:1996} D'Hoker, Krichever, and Phong 
derived the prepotential for the $\cN=2$ $\U(N)$ theory with $N_f$ flavors 
from a Seiberg-Witten curve of the form\footnote{Note:  
$\Lambda^{2N-N_f}$ in ref.~\cite{D'Hoker:1996} 
differs from ours by a factor of 4,  except in eq. (4.34).
In the e-print version of ref.~\cite{D'Hoker:1996}
the factor of 4 in eq. (2.6) should be omitted,
and the right hand sides in eq. (2.8) should be multiplied by $1/4$.
These typos are corrected in the published version.}
\beq
\label{SWcurve}
y^2 = \left[  \prodN (x-e_i)  + 4 \Lambda^{2N-N_f} T(x) \right]^2
- 4 \Lambda^{2N-N_f} \prod_{I=1}^{N_f}(x+m_I) \,.
\eeq 
In their work  
the $(N_f -N)$th order polynomial $T(x)$ was left unspecified
(although two different 
candidates \cite{Hanany:1995}, \cite{Krichever:1996} were presented) 
since, as shown in sec.~II.c of that paper, 
the prepotential $\cF(a)$ is independent of $T(x)$.  
This is because $T(x)$ can always be absorbed into a redefinition
of the $e_i$,
and $\cF(a)$ is insensitive to a redefinition of $e_i$.
However, since $T(x)$ is tied to the definition of $e_i$,
its form will affect the relation between $a_i$ and $e_i$.

Our matrix model calculation  of the relation between $a_i$ and $e_i$
\eqs{ae} implies (via eq.~(3.10) of ref.~\cite{D'Hoker:1996})
a specific form for $T(x)$, namely 
\beq
\label{Tform}
T(x) =   \half \poly{x} + \cO( \Lambda^{2N-N_f} )
       = \half \! \sum_{k=0}^{N_f-N} \tilde{t}_k x^{N_f-N -k}
                      + \cO( \Lambda^{2N-N_f} )\,,
\eeq
and thus corresponds to a specific choice of the $e_i$.
(Our perturbative matrix model calculation only yields 
a result valid to one-instanton accuracy.)
The Seiberg-Witten curve \eqs{SWcurve} corresponding to
eq.~\eqs{Tform} has the form 
\bea
\label{ourcurve}
y^2 &=& \prodN (x-e_i)^2 - f(x) \,, 
\non\\
f(x) &=& 4 \Lambda^{2N-N_f} 
\left( \prod_{I=1}^{N_f} (x+m_I) - \poly{x} \prodN (x-e_i)  \right)
+ \cO( \Lambda^{4N-2N_f}  )\,.
\eea
The definition of $\poly{x}$,
given below eq.~\eqs{polydef},
ensures that $f(x)$ is at most an $(N-1)$th order polynomial.
Thus, the choice of $e_i$ in the matrix model is such that
none of the coefficients of $x^N$ or higher powers 
in $y^2$ receive $O(\Lambda^{2N-N_f} )$ corrections.
(However, as we discuss below, the gauge-invariants $\vev{u_n}$ 
do receive corrections.)
As we will see in the next section, 
this is exactly what the saddle-point solution of the matrix model
requires.

It is curious to note that the form of $T(x)$ 
proposed in ref.~\cite{Krichever:1996} 
and on the right hand side of eq. (2.8) in ref.~\cite{D'Hoker:1996} 
is\footnote{Taking into account the correction in the previous footnote.}
$T(x) = {1\over 4} \sum_{k=0}^{N_f-N} \tilde{t}_k x^{N_f-N -k}$,
precisely one-half of that in eq.~\eqs{Tform}.
Why the difference?   

Consider the gauge-invariant variables
$\vev{u_n} = {1 \over n} \vev{\tr (\phi^n) }$,
which classically have the values $(u_n)_{\rm cl} = (1/n) \sumN e_i^n$.
Quantum mechanically, these may be computed 
via \cite{Dijkgraaf:2002d, Naculich:2002}
$ \vev{u_n} = (1/2 \pi i n) \sumN \oint_{A_i}  x^{n-1} \laSW$,
where $\laSW$ is the Seiberg-Witten differential.
They may also be computed in the matrix model \cite{Naculich:2002},
starting from the correlators $\vev{\tr( \Phi^n)}$
(and modifying the expressions of ref.~\cite{Naculich:2002} 
to include the $\vev{\tr( \Phi^n)} \vert_{\rm disk} $ contribution,
as in eq.~\eqs{amat} of this paper; see sec.~8).
It is easily shown that for $N_f < N$ (in which case $T(x)$ vanishes)
$\vev{u_n} = (u_n)_{\rm cl}$ for 
$n=1, \cdots, N$ \cite{Cachazo:2002,Naculich:2002}.
When $N_f \ge N$, however, $\vev{u_n}$ with $2N- N_f \le n \le N$
can get $\cO(\La^{2N-N_f})$ corrections. 

As stated above, choosing a particular $T(x)$ corresponds
to a particular choice of parameters $e_i$ 
used to parametrize the moduli space.   
It is possible to define the $N$ parameters $e_i$ 
so that the relation $\vev{u_n} = (1/n) \sumN e_i^n$ 
continues to hold quantum mechanically for $n=1, \cdots, N$.
This requirement then leads to the form of $T(x)$ 
in ref.~\cite{Krichever:1996} \cite{D'Hoker:1996}
(see however ref.~\cite{Slater:1997}).
In contrast, for the choice of $T(x)$ in eq.~\eqs{Tform},
$\vev{u_n} = (u_n)_{\rm cl}$ no longer holds
at the one-instanton level.

\setcounter{equation}{0}
\section{Matrix model derivation of the Seiberg-Witten curve}

In this section, we will derive the form of the Seiberg-Witten
curve for $\cN=2$ U($N$) gauge theory with $N_f < 2N$ fundamental
hypermultiplets by solving the matrix model integral
using saddle-point methods (for a review of this method, 
see, e.g., ref.~\cite{DiFrancesco:1993}).

Our starting point is the  matrix model partition function \eqs{partition}
\beq
Z = {1\over \mathrm{vol}(G)} 
\int \D\Phi \, \D Q^I \D \tQ_I 
\exp 
\left( - \frac{1}{\gs} 
 W_0(\Phi) 
- {1 \over \gs} 
\sumI \left[ \tQ_I \phi\, Q^I + m_I \tilde{Q}_I Q^I  \right]
 \right) \,.
\eeq
Diagonalizing $\Phi$ 
and integrating over $Q$, $\tQ$, 
one obtains ($\lam_i$ are the eigenvalues of $\Phi$)
\cite{Dijkgraaf:2002a, McGreevy:2002}
\beq
\label{part}
Z \propto 
\int \prod_{i=1}^M  \D \lam_i 
\exp \left( - \frac{1}{\gs}
\sum_i  W_0 (\lam_i) 
+  2 \sum_{i < j} \log (\lam_i - \lam_j)
- \sumI  \sum_i \log  (\lam_i  + m_I) 
\right) 
\eeq
The saddle-point equation is obtained by varying the action with respect
to $\lam_i$:
\beq	\label{noforce}
- {1\over \gs} \Wp (\lam_i) + 2 \sum_{j \not=i} {1 \over \lam_i -\lam_j} 
-  \sumI {1 \over \lam_i + m_I }
= 0\,.
\eeq
To solve (\ref{noforce}), it is standard procedure \cite{DiFrancesco:1993}
to introduce the trace of the resolvent 
\beq
\label{resolvedef}
\om(x) = \frac{1}{M}\tr \left(\frac{1}{\Phi-x}\right) 
= {1\over M} \sum_i {1\over \lam_i-x} 
\eeq
which can be shown to satisfy \cite{DiFrancesco:1993}
\bea
\label{resolveqn}
&&\om^2(x)+ {\Wp (x) \over \gs M} \, \om(x)
 + {1\over \gs M^2} \sum_i {\Wp (x) - \Wp (\lam_i) \over x - \lam_i} \\ \non
&& - {1\over M} \om'(x)  
+ {1\over M^2} \sum_i \sumI {1\over (\lam_i-x)(\lam_i + m_I)}
=0 \,.
\eea 

Now we let $\gs \rar 0$, $M \to \infty$, with $S= \gs M$ held fixed.
We also hold $N_f$ fixed;
in this, our approach differs from ref.~\cite{McGreevy:2002}.
In this limit, the last two terms of eq.~\eqs{resolveqn} vanish.

The large-$M$ limit expressions are conveniently written in 
terms of the density of eigenvalues
\beq
\rho(\lam)= {1\over M} \sum_i \delta(\lam-\lam_i) \,, \qquad\qquad
\int \rho(\lam)\, \D \lam=1 \,.
\eeq
In this language the resolvent becomes
\beq
\label{resolvent}
\om(x) = \int \D \lam {\rho(\lam) \over \lam - x} \,,
\qquad\qquad
\rho(\lam)=
{1 \over 2 \pi i} \left[ \om(\lam + i \eps) - \om (\lam-i\eps) \right]
\eeq
and eq.~(\ref{resolveqn}) can be rewritten as
\beq
\label{resolventeq}
\om^2(x) + {\Wp (x) \over S} \om(x)
 + {1\over 4 S^2 } f(x) = 0
\eeq
where
\beq
\label{fdef}
f(x) = 4 S  \int  \D \lam  \, \rho(\lam) \,  
{W_0'(x) - W_0'(\lam) \over x - \lam}
\eeq
is an (as yet) arbitrary  $(N-1)$th order polynomial. 
Defining 
\beq
\label{ydef}
y(x)  = 2S \om(x) + \Wp (x)
\eeq
one may rewrite eq.~\eqs{resolventeq} as
\beq
\label{hyper}
y^2 =  \Wp (x)^2 - f(x) , \qquad f(x) = \sum_{n=0}^{N-1} b_n x^n\,.
\eeq 
This equation characterizes a hyperelliptic Riemann surface.
When the roots of $\Wp (x)$ are well-separated and 
$f(x)$ is a small correction to $\Wp (x)$,
the curve has $N$ cuts in the $x$ plane, 
centered approximately on the roots of $\Wp (x)$.
The eigenvalues of $\Phi$ are clustered along these cuts.
The function $f(x)$ determines the distribution of the 
eigenvalues of $\Phi$ among the $N$ cuts, and the spreading
of those eigenvalues due to eigenvalue repulsion. 
Let $M_i$ denote the number of eigenvalues along the $\ith$:
\beq
\label{icut}
M_i = M  \int_{i} \D \lam\, \rho(\lam) \,.
\eeq
Define $S_i =\gs M_i$, which remains finite in the $M$, $M_i \to \infty$
limit.
Then, using  eqs.~\eqs{resolvent} and \eqs{ydef}, 
we see that eq.~\eqs{icut} may be rewritten
\beq
\label{Aperiod}
S_i = \, - \,{1 \over 4 \pi i} \oint_{A_i} y \, \D x 
\eeq
where $A_i$ denotes the contour surrounding the $i$th cut.
This is eq.~(3.10) of \cite{Dijkgraaf:2002a} (up to a factor of 2;
the sign depends on the direction of the contour integrals,
which we take to be counterclockwise).
Up to this point, we have just been following ref.~\cite{Dijkgraaf:2002a}.

As in ref.~\cite{Cachazo:2002}, 
we denote by $P$ and $Q$ the points $x = \infty$ 
on the two sheets of the curve \eqs{hyper}.
(If one needs a cutoff for an integral, 
one takes $P$ and $Q$ to be at $x=\Lambda_0$ with $\Lambda_0$ large.)
To be specific, let $P$ be on the sheet on which $\Wp (x) - y(x)$ goes
to zero as $x \to \infty$.
Also, denote by $C_i$ a path from $Q$ to $P$ 
that passes through the $\ith$. 
The Riemann surface of genus $N-1$ described by the curve \eqs{hyper}
can be given a canonical homology basis as follows:
$A_i$ ($i=1, \ldots, N-1$) and $B_i = C_i - C_N$ ($i=1, \ldots, N-1$).

Our goal in the remainder of this section is to use matrix-model methods 
to determine the explicit form of $f(x)$ in the spectral curve
\eqs{hyper}.
This will in turn yield the (hyperelliptic) Seiberg-Witten curve
for the U($N$) theory with $N_f$ fundamental hypermultiplets.
The saddle-point evaluation of the partition function \eqs{part} gives 
(here we need to keep the first subleading term since it contributes to 
$\fdis$)
\bea
Z 
&=&  
\exp \Bigg(
- {S\over \gs^2}  \int\D\lam \, \rho(\lam) \, W_0(\lam) 
+ {S^2\over \gs^2}  \int \D\lam \, \D\lam' \, \rho(\lam)\, \rho(\lam') 
\log( \lam - \lam') 
\non\\ && 
\qquad\;\; 
- {S \over \gs} \sumI \int\D\lam \, \rho(\lam) \, \log (\lam + m_I) \Bigg)
\eea
{}from which we infer
\beq \label{freM}
\fre = - S \int \D\lam \, \rho(\lam) \, W_0(\lam) 
+ S^2  \int \D\lam \, \D \lam' \, \rho(\lam) \, \rho(\lam') 
\, \log( \lam - \lam') 
\eeq
and
\beq
\fdis = -S \sumI \int\D\lam \, \rho(\lam) \, \log (\lam + m_I) \,.
\eeq
In order to compute $\Weff$, we need the variation of 
$\fre$ under a small change in $S_i$. 
{}From (\ref{icut}) we see that such a variation can be implemented by letting 
$\rho(\lam) \to \rho(\lam) + (\delS_i/S) \delta(\lam -  e_i)$
where $e_i$ refers to an arbitrary, but fixed, point along the $\ith$.
Using this result in (\ref{freM}) gives\footnote{See \cite{Dijkgraaf:2002a} 
and  appendix A of ref.~\cite{Ferrari:2002} for related discussions.}
\bea
\delta \fre
= \delS_i \left[ - W_0(e_i) 
+ 2 S \int \D \lam \, \rho (\lam) \, \log (\lam-e_i) \right]\,.
\eea
This may be rewritten as
(here const refers to a constant of integration)
\bea
{\partial \fre \over \partial S_i}
&=& \int_{e_i}^P \D x ~ \Wp (x) 
- 2 S \int \D \lam \, \rho(\lam)  \int_{e_i}^P {\D x\over x- \lam} 
+ \con
\non\\
&=& \int_{e_i}^P \D x \left( \Wp (x) +  2 S \, \om(x) \right)  + \con \non\\
&=&  \int_{e_i}^P  y \, \D x  + \con 
\eea
which is just eq. (3.11) of \cite{Dijkgraaf:2002a}.
Using the fact that $y$ differs only by a sign on the two sheets,
together with the definition $B_i= C_i - C_N$,
we may rewrite this as
\bea
\label{freeresult}
{\partial \fre \over \partial S_i}
&=& \half \int_{Q}^{e_i} y \, \D x  + \half \int_{e_i}^P y \, \D x 
+ \con \non\\
&=& \half \int_{C_i}  y \, \D x + \con \non\\
&=& \half \int_{B_i}  y \, \D x+ \half \int_{C_N}  y \, \D x  + \con\,.
\eea
For $\Weff$, we will also need 
\bea
\label{fdiskresult}
\fdisk
&=&  -S \sumI 
\int \D\lam \, \rho(\lam) \int_{-m_I}^P {\D x\over \lam-x} + \con \non\\
&=& -S \sumI \int_{-m_I}^P \om(x) \,\D x + \con \non\\
&=& -\half \sumI \int_{-m_I}^P  y(x) \, \D x + \con
\eea
where we absorb the $S_i$-independent $ W_0(P)-W_0(-m_I)$ terms 
into the integration constant.

We now use eqs.~\eqs{freeresult} and \eqs{fdiskresult}
in the effective superpotential (setting $N_i=1$)
\bea
\label{Weffy}
\! \Weff \!&=& \!\!- \sumN  {\partial \fre \over \partial S_i} - \fdis
  +  2 \pi i \tauo \sumN  S_i \non\\
&=& \!\! - \half \sum_{i=1}^{N-1} \oint_{B_i} \! y \, \D x - 
\half N \int_{C_N} \!\! y\, \D x
+ \half \sumI \int_{-m_I}^P \!\! y \, \D x
- \half \tauo \sumN \oint_{A_i} \!\! y \, \D x + \con .
\eea
In the prescription relating the matrix model and the $\cN=2$ gauge theory 
we are instructed to extremize $\Weff$ with respect to $S_i$.
Since the $S_i$'s are determined by $f(x)$ and therefore
by the $b_n$'s through eqs.~\eqs{hyper} and \eqs{Aperiod},
we may equivalently vary \eqs{Weffy} with respect to 
$b_n$ \cite{Cachazo:2002}. 
{}From eq.~\eqs{hyper}, one sees that
$(\partial y /\partial b_n)\D x = - \half x^n \D x/y$.
For $0 \le n \le N-2$, these form a complete basis of holomorphic
differentials on the Riemann surface \cite{Farkas:1991}.
We may therefore change basis to the unique basis of 
holomorphic differentials $\zeta_k$ dual to the 
homology basis, i.e.,  $\oint_{A_i} \zeta_k = \delta_{ik}$.
Consequently, the equations $\delta \Weff/ \delta b_n = 0$ 
for $0 \le n \le N-2$ may be rewritten
\beq
0 =  - \sum_{i=1}^{N-1} \oint_{B_i} \zeta_k - N \int_Q^P \zeta_k
      +  \sumI \int_{-m_I}^P \zeta_k
\eeq
where $\sumN \oint_{A_i} \zeta_k= 0$ because the sum of $A_i$ cycles
is a trivial cycle.
The first term just yields $\sum_{i=1}^{N-1} \tau_{ik}$,
which is an element of the period lattice.
Hence\footnote{This equation was obtained in ref.~\cite{Cachazo:2002}
for the case $N_f=0$ by a somewhat different approach. Here we have 
derived it using only matrix-model methods.}
\bea \label{Abel}
&& N \int_P^Q \zeta_k +  \sumI \int_{-m_I}^P \zeta_k = \non \\
&& N \int_{p_0}^Q \zeta_k - (N-N_f) \int_{p_0}^P \zeta_k 
- \sumI \int_{p_0}^{-m_I} \zeta_k =
0 \quad
{\rm (modulo~the~period~lattice)}
\eea
where $p_0$ is an arbitrary (generic) point on the Riemann surface. 
It now follows from Abel's theorem \cite{Farkas:1991}
that there exists a function $\psi(x)$
on the Riemann surface with an $N$th order pole at $Q$, 
an $(N-N_f)$th order zero (or pole, if $N_f > N$) 
at $P$,
and simple zeros at $-m_I$ for $I=1, \ldots, N_f$. 
As we will now show, this requirement suffices to fix the
form of $f(x)$, and therefore the Seiberg-Witten curve.

For $0 \le N_f < N$, the function
$\psi(x)$ is simply (proportional to) the resolvent:
\beq \label{psiI}
\psi(x) = y - \Wp(x) = \sqrt{\Wp(x)^2 - f(x)}  - \Wp(x) , \qquad 0 \le N_f < N
\eeq
This can be seen as follows:  
$\psi(x)$ has an $N$th order pole at $Q$,
and (at least) a simple zero at $P$ 
(because $f(x)$ is a polynomial of at most $(N-1)$th order).
Abel's theorem yields $N-1$ conditions and 
therefore completely constrains the remaining zeros.
Thus $\psi(x)$ must have a simple zero at $x=-m_I$, 
so $f(x)$ must contain a factor $(x+m_I)$  for each $I$.
For $\psi(x)$ to have an $(N-N_f)$th order zero at $P$,
$f(x)$ can be of $N_f$th order at most.
These two conditions require 
$f(x)  \propto  \prod_{I=1}^{N_f} (x+m_I) $.
Naming the constant of proportionality $4 \La^{2N-N_f}$,
and setting $\alpha=1$ in eq.~\eqs{W},
we see that the spectral curve \eqs{hyper} is given by
\beq
\label{curveI}
y^2 = \prodN (x-e_i)^2 - 4 \La^{2N-N_f} \prod_{I=1}^{N_f} (x+m_I) 
\eeq
precisely the Seiberg-Witten curve \cite{Seiberg:1994}--\cite{Krichever:1996}
for $N_f<N$.
(It should also be possible to determine this constant of proportionality
by setting $\delta \Weff/ \delta b_{N-1} = 0$,
and using the gauge theory relation 
$2 \pi i \tau(\Lambda_0) = (2N- N_f) \log (\Lambda/\Lambda_0)$
and the fact that \cite{Cachazo:2002}
$\sumN \oint_{A_i} \!\! y \, \D x  = - \pi i b_{N-1}$.)

For $N \le N_f  < 2N$, the function
$\psi(x)$ is not given by the resolvent
but by a related function
\beq \label{psiII}
\psi(x) = \sqrt{ A(x)^2 - g(x) } - A(x),  \qquad N \le N_f < 2N
\eeq
where $A(x)$ is an $N$th order polynomial
and $g(x) \propto \prod_{I=1}^{N_f} (x+m_I)$. 
(As before, we name the proportionality constant $4 \La^{2N-N_f}$.)
Under these conditions, $\psi(x)$ vanishes at $x=-m_I$,
for $I=1, \ldots N_f$, has an $N$th order pole at $Q$, 
and an $(N_f-N)$th order pole at $P$. 
For $\psi(x)$ to be a function on the Riemann surface \eqs{hyper}, 
the square root in $\psi(x)$ must be proportional to $y(x)$,   
that is (normalizing appropriately)
\beq
A(x)^2 - 4 \La^{2N-N_f} \prod_{I=1}^{N_f} (x+m_I) 
= \Wp(x)^2 - f(x)
\eeq
where $f(x)$ is a polynomial of order at most $(N-1)$.
The solution to this, to $\cO(\La^{2N-N_f})$,  is 
\bea
A(x) &=& \prodN (x-e_i) + 2 \La^{2N-N_f} \poly{x},  \non\\
f(x) &=& 4 \La^{2N-N_f} 
\left( \prod_{I=1}^{N_f} (x+m_I) - \poly{x} \prodN (x-e_i)   \right)
\eea
where $\poly{x}$ is defined below eq.~\eqs{polydef},
and again we have set $\alpha=1$ in eq.~\eqs{W}.
Thus the spectral curve \eqs{hyper} and function $\psi(x)$ are given by
\bea
\label{psiIII}
y^2 & = & \prodN (x-e_i)^2 - f(x), \qquad N \le N_f < 2N \non\\
\psi(x) &=& y - A(x), 
\eea
in agreement with the Seiberg-Witten curve for $N \le N_f < 2N$
\cite{Seiberg:1994}--\cite{Krichever:1996}
but with a particular choice of subleading term $\poly{x}$.
(This form of the curve was already obtained \eqs{ourcurve}
in the previous section by comparing our perturbative
matrix model calculation with the curve in ref.~\cite{D'Hoker:1996}.
The subleading term $\poly{x}$ simply corresponds to 
a particular choice of moduli parameters $e_i$ 
picked out by the matrix model.)

Thus, for both $N_f < N$ and $N \le N_f < 2N$,  
the spectral curve obtained from the matrix-model 
saddle-point integral agrees precisely with the known
Seiberg-Witten curve \eqs{SWcurve} 
for the $\cN=2$ U($N$) gauge theory with $N_f$ 
fundamental hypermultiplets.  

Finally, from the properties of $\psi(x)$ 
(\ref{psiI}), (\ref{psiII}), 
we see that
\beq \label{h}
h(x) \D x =   {\D \psi \over \psi}
\eeq
is a meromorphic differential with simple poles 
at $P$, $Q$, and $x=-m_I$ and residues
$N-N_f$, $-N$, and $1$  respectively.
These conditions imply that the meromorphic 
differential given by $\laSW = x\, h(x) \D x$ 
has all the correct properties to be the 
Seiberg-Witten differential \cite{Seiberg:1994,Krichever:1996,Krichever:1997}. 
Moreover, using the specific forms of $\psi(x)$ given in 
eqs.~\eqs{curveI} and \eqs{psiIII},
we obtain exactly the form of the $\laSW$ given in ref.~\cite{D'Hoker:1996}.

\setcounter{equation}{0}
\section{Derivation of the Seiberg-Witten differential}
In the previous section we obtained an expression \eqs{h} 
related to the Seiberg-Witten differential $\laSW$. 
Although this form can be motivated from the 
Calabi-Yau approach \cite{Cachazo:2002,Ookouchi:2002} 
it does not constitute a genuine 
matrix-model derivation of $\laSW$.
In this section we present a derivation of $\laSW$
entirely within the framework of the matrix model. 

In the Seiberg-Witten approach, the gauge-theory expectation 
value of $\tr\,\phi^n$ is calculated via \cite{Dijkgraaf:2002d, Naculich:2002}
\be \label{vevsw}
\vev{\tr\,\phi^n} = \frac{1}{2\pi i} \sumN \oint_{A_i}  x^{n-1} \la_{SW} \,.
\ee
The relation between the gauge theory vev and matrix model
quantities is
\be \label{vevmm}
\vev{\tr\,\phi^n} =  \left[ \sumjN {\pa \over \pa S_j} \gs \vev{\tr \Phi^n}_{\mathrm{sphere}} 
+ \vev{\tr\,\Phi^n}_{\mathrm{disk}} \right] \vevS
\ee
which generalizes eq.~(5.10) in ref.~\cite{Naculich:2002} 
to the case when boundaries are present (see also \cite{Gopakumar:2002}).
The  derivation of eq.~\eqs{vevmm} is similar to that of 
eq.~\eqs{amat} of this paper but uses the deformation
$\tW(\Phi,Q,\tQ) =  W(\Phi,Q,\tQ) + \eps \,(1/n)\, \tr\, (\Phi^n) $. 

The matrix-model expectation values $\vev{\tr\, \Phi^n}$ in eq.~\eqs{vevmm}
may be expressed in terms of the 
resolvent \eqs{resolvedef}
\bea
\label{generating}
\om(x) &=& \,- \,\frac{1}{M}\left\langle \tr\frac{1}{x-\Phi}\right\rangle = 
\, -\, \frac{1}{M} \sum_{n=0}^{\infty}x^{-n-1} \vev{\tr\, \Phi^n} \non\\
\vev{\tr\, \Phi^n} &=&-\,  \frac{M}{2\pi i} \sumN \oint_{A_i}  x^n 
\,\om(x)\, \D x
\eea
which acts as a generating function for the expectation values.
To proceed, we rewrite the last term in (\ref{resolveqn}) as
\bea
{1\over M^2} \sum_i \sumI {1\over (\lam_i-x)(\lam_i + m_I)}
&=&
{1\over M^2} \sum_i {1\over \lam_i-x} \sumI {1 \over x+m_I}
\,-\, {1\over M^2} \sumI {1\over x+m_I} \sum_i {1\over \lam_i + m_I}
\non\\
&=& {1\over M} \sumI { \om (x) - \om (-m_I) \over x+m_I} 
\eea 
so that eq.~\eqs{resolveqn} becomes
\bea
\label{loopeqn}
&&
\om^2(x)+ {\Wp (x) \over \gs M} \, \om(x)
 + {1\over \gs M^2} \sum_i {\Wp (x) - \Wp (\lam_i) \over x - \lam_i}
 \\ \non && 
- {1\over M} \om'(x)  
+ {1\over M} \sumI { \om (x) - \om (-m_I) \over x+m_I} 
=0 \,.
\eea 
Next, we expand $\om(x)$ as  
\be \label{omexp}
\om(x) = \sum_{\chi\le 2} \frac{1}{M^{2-\chi}} \om_{1-\chi/2}(x) = 
\om_0(x) + \frac{1}{M} \om_{1/2}(x) + \cO(\frac{1}{M^2})\,.
\ee
Using the method developed in 
ref.~\cite{Ambjorn:1992}\footnote{See also the recent 
paper \cite{Ashok:2002}.}, 
we can solve the loop-equation (\ref{loopeqn}) 
order-by-order in $1/M$,
which in principle will give us $\vev{\tr\, \Phi^n}$ 
to arbitrary order in the topological expansion.
For eq.~\eqs{vevmm}, however,
we will only need $\om_{\rm s}(x) \equiv \om_0(x)$ and 
$\om_{\rm d}(x) \equiv \om_{1/2}(x)$. 
Inserting \eqs{omexp} into eq.~\eqs{loopeqn}, 
and using the fact \cite{Ambjorn:1992}\footnote{The 
relation to the formul\ae{} in ref.~\cite{Ambjorn:1992} is: 
$\frac{1}{M}\om'(x) = -\frac{1}{M^2} \langle 
\tr \frac{1}{x-\Phi} \tr \frac{1}{x -\Phi} \rangle_{\mathrm{conn}}
$.}
that the $(1/M) \om'(x)$ term is $\cO(1/M^2)$,
we find 
\bea
\label{omhalf}
\om_{\rm s}(x) &=&  {1 \over 2 S} \left[ y - \Wp(x) \right]\,,
\non\\
\om_{\rm d}(x) &=& -\frac{S}{y} \sumI \frac{ \om_{\rm s}(x)
- \om_{\rm s}(-m_I)}{x+m_I} \,,
\eea
where $y^2 = W_0'(x)^2 - f(x)$. 
This result, together with \eqs{generating},
allows us to write the contributions to $\vev{\tr\, \Phi^n}$ 
at the sphere ($\chi=2$) and disk ($\chi=1$) levels as 
\bea 
\label{omint}
\vev{\tr\, \Phi^n}_{\mathrm{sphere}} 
&=&-\,  \frac{M}{2\pi i} \sumN \oint_{A_i} x^n \,\om_{\rm s}(x)\, \D x \,,
\non \\
\vev{\tr\, \Phi^n}_{\mathrm{disk}} 
&=&-\,  \frac{1}{2\pi i} \sumN \oint_{A_i} x^n 
\,\om_{\rm d}(x) \, \D x \,.
\eea
Inserting these expressions into eq.~(\ref{vevmm}) 
and comparing with (\ref{vevsw}) 
one can read off
\be
\la_{SW} = x \left[ \sumN \frac{\pa}{\pa S_i} (-S \om_{\rm s}(x) ) 
- \om_{\rm d}(x)\right]\vevS \D x \,.
\ee
This generalizes eq.~(5.3)
in v3 of ref.~\cite{Gopakumar:2002}
to the case when boundaries are present.

Using eq.~\eqs{omhalf}, we have
\be
\sumN \frac{\pa}{\pa S_i} (-S \om_{\rm s}(x)) 
= \,-\, \half \sumN \frac{\pa y}{\pa S_i} \,.
\ee
This expression has unit $A_i$-periods,
\be \label{unitperiods}
\frac{1}{2\pi i} \oint_{A_i} 
\left[ -\, \half \sum_{j=1}^N \frac{\pa y}{\pa S_j} \right] \, \D x
=  \sum_{j=1}^N \frac{\pa}{\pa S_j} 
\left[ -\, \frac{1}{4 \pi i}\oint_{A_i} y\,  \D x \right]
= \sum_{j=1}^N \frac{\pa}{\pa S_j} S_i = 1
\ee
using the definition of $S_i$ (\ref{Aperiod}).
Moreover, by writing ($b_n$ was defined in eq.~(\ref{hyper}) )
\bea \label{thirdkind}
\,-\, \half \sumN \frac{\pa y}{\pa S_i} 
= \,- \, \half \sumN \sum_{n=0}^{N-1} \frac{\pa b_n}{\pa S_i} \frac{\pa y}{\pa b_n}
&=& \,- \, \half  
\frac{\pa y}{\pa b_{N-1}} \sumN \frac{\pa b_{N-1}}{\pa S_i} + 
{\rm holomorphic} \non \\
&=& {N x^{N-1} \over y}+ {\rm holomorphic}
\eea
we see that this expression has simple poles at $P$ and $Q$ with residues
$\pm N$, and no other poles.
The properties \eqs{unitperiods} and \eqs{thirdkind} suffice to show that 
\be
\,-\, \half \sumN \frac{\pa y}{\pa S_i} =  \frac{W_0''(x)}{y}
\ee
as the function on the r.h.s. has the same properties.

To simplify the remainder of the discussion, we consider $N_f<N$.
In this case, we found in the previous section that
$f(x)  \propto  \prod_{I=1}^{N_f} (x+m_I) $, so $f(-m_I)=0$.
The contours in eq.~(\ref{omint}) are on the sheet on which
$y = +\Wp(x) + \ldots$, and on this sheet, 
eqs.~\eqs{ydef} and \eqs{hyper}  imply $\om_{\rm s}(-m_I)=0$
so this term drops out of eq.~(\ref{omhalf}),
yielding
\be
\om_{\rm d}(x) = \,-\, \frac{y - \Wp(x)}{2y} \sumI \frac{1}{x+m_I} 
             = \,-\, \frac{y - \Wp(x)}{2y} {f' \over f}
\ee
Collecting the above results one finds
\be
\la_{SW} = \frac{x}{y}
\left[ W^{\prime\prime}_0 (x) - \half (\Wp(x)  - y) \frac{f'}{f}\right]
\ee
which is in perfect agreement 
with the $N_f<N$ result in ref.~\cite{D'Hoker:1996}.

\setcounter{equation}{0}
\section{Summary}
In this paper we have continued the program initiated in \cite{Naculich:2002} 
for analyzing $\cN=2$ gauge theories within the matrix model 
approach \cite{Dijkgraaf:2002a}--\cite{Dijkgraaf:2002d}; here we included 
matter in the fundamental representation of $\U(N)$. This addition 
exposes new features of the method, one of which is the appearance 
of disk diagrams that contribute to the free energy. 
Similarly, the tadpole diagrams necessary for 
computing the periods $a_i$ also have a contribution from 
disk diagrams.  
We computed the relation between $a_i$ and the classical moduli $e_i$,
as well as the $\cN=2$ prepotential $\cF(a)$, finding
complete agreement with known results.

An interesting feature of our calculation is that the two cases 
$N_f < N$ and $N \le N_f < 2N$ are on the same footing and can 
be treated using the same method within the matrix model approach. 
The only difference between the two cases is 
the appearance of the polynomial $\poly{x}$ when $N_f\ge N$, cf. (\ref{ae}). 
In the final expression for the prepotential, however,
$\poly{x}$ disappears. 
In section \ref{sT} we discussed the meaning of $\poly{x}$,
explaining how it affects the form of the 
Seiberg-Witten curve when $N_f \ge N$. 

{}From the point of view of computational efficiency, 
the matrix model approach cannot, in its present form, 
compete with other methods for computing
multi-instanton contributions 
\cite{Chan:1999}--\cite{Nekrasov:2002}. 
However, it would be interesting to connect these approaches
with the matrix model perspective to improve our 
understanding of multi-instanton effects. 

In sections 7 and 8 we presented derivations, 
entirely within the context 
of the matrix model, of the Seiberg-Witten curve 
and differential 
for the $\cN=2$ $\U(N)$ theory with $N_f<2N$ flavors. 
The contribution to the free energy from disk diagrams (\ref{fdiskresult})
played an important role in the analysis. 
A comparison of (\ref{psiI}) and (\ref{psiIII})  
exhibits the difference between the Seiberg-Witten curves
for $N_f<N$ and $N\le N_f < 2N$. 
In the latter case, the matrix model makes a specific choice 
for the modification of the curve.
This result was also inferred in sec.~6 from the perturbative
calculation.

\section*{Acknowledgments}
We would like to thank J. McGreevy for conversations 
and R. Gopakumar for correspondence. 
HJS would like to thank the string theory group and Physics 
Department of Harvard University for their hospitality extended 
over a long period of time.

\appendix

\end{document}